\documentclass{actasen}

\begin{document}

%%ÉèÖÃÊ×Ò³Ò³Âë
\setcounter{page}{115}

\Volume{2011}{35}% Äê¡¢¾í

%%ҳüÉèÖÃ

\runheading{ZHANG Zhao-wen}%

\title{Self-consistent Thomas-Fermi Approximation for Equation of State at Subnuclear Densities$^{\dag}$}

\footnotetext{$^{\dag}$ Supported by the National Natural Science Foundation of China(Grants No. 11075082 and No. 11375089)

Received 2013--05--02; revised version 2013--08--06 \\
%$^{\star}$ A translation of {\it Acta Astron. Sin.~}
%Vol. 55, No. 2, pp. 116--126, 2014 \\
\hspace*{5mm}$^{\bigtriangleup}$ songtc@nankai.edu.cn\\

\noindent 0275-1062/01/\$-see front matter $\copyright$ 2011 Elsevier
Science B. V. All rights reserved. %%

\noindent PII: }

\enauthor{ZHANG Zhao-wen, SHEN Hong$^{\bigtriangleup}$ }{School of Physics, Nankai University, Tianjin 300071}

\abstract{The self-consistent Thomas--Fermi approximation is an essential method for studying the non-uniform nuclear matter with relativistic mean-field theory. In this method, the nucleon distribution in the Wigner--Seitz cell is obtained self-consistently. We make a detailed comparison between the present results and previous calculations in the Thomas--Fermi approximation with a parameterized nucleon distribution that has been adopted in the widely used Shen equation of state.}


\maketitle

\section{Introduction}

The equation of state (EOS) of dense matter plays a very important role in studying many astrophysical phenomena such as supernova explosions and neutron star formations\rf{1,2,3,4}. One of the most commonly used EOS in supernova simulations is called the Shen EOS\rf{4,5,6}, which used a relativistic mean-field (RMF) model and the parameterized Thomas--Fermi (PTF) approximation. In this paper, we use the self-consistent Thomas--Fermi (STF) approximation to study the non-uniform nuclear matter, and then examine the differences between the STF and PTF approximations. We also consider the effect of the bubble phase which may appear before the transition to uniform matter.

We use the RMF theory for the effective nuclear interaction, in which nucleons interact via the exchange of isoscalar scalar and vector mesons ($\sigma$ and $\omega$) and an isovector vector meson ($\rho$). In the present work, we employ the RMF theory with the parameter set TM1\rf{7}.

\section{Formalism}

We first give a brief description of the RMF theory. For a system consisting of protons, neutrons, and electrons, the Lagrangian density reads
\begin{small}
\begin{eqnarray}
\label{eq:LRMF}
{\cal L}_{\rm{RMF}} & = & \sum_{i=p,n}\bar{\psi}_i\left[i\gamma_{\mu}\partial^{\mu} -M
-g_{\sigma}\sigma-g_{\omega}\gamma_{\mu}\omega^{\mu}
-g_{\rho}\gamma_{\mu}\tau_a\rho^{a\mu}
-e \gamma_{\mu}\frac{1+\tau_3}{2} A^{\mu}
\right]\psi_i  \nonumber\\
& & +\bar{\psi}_{e}\left[i\gamma_{\mu}\partial^{\mu} -m_{e}
+e \gamma_{\mu} A^{\mu} \right]\psi_{e}  \nonumber\\
 && +\frac{1}{2}\partial_{\mu}\sigma\partial^{\mu}\sigma
-\frac{1}{2}m^2_{\sigma}\sigma^2-\frac{1}{3}g_{2}\sigma^{3}
-\frac{1}{4}g_{3}\sigma^{4} \nonumber\\
 && -\frac{1}{4}W_{\mu\nu}W^{\mu\nu}
+\frac{1}{2}m^2_{\omega}\omega_{\mu}\omega^{\mu}
+\frac{1}{4}c_{3}\left(\omega_{\mu}\omega^{\mu}\right)^2   \nonumber\\
 && -\frac{1}{4}R^a_{\mu\nu}R^{a\mu\nu}
+\frac{1}{2}m^2_{\rho}\rho^a_{\mu}\rho^{a\mu}
-\frac{1}{4}F_{\mu\nu}F^{\mu\nu},
\end{eqnarray}
\end{small}
where $W^{\mu\nu}$, $R^{a\mu\nu}$, and $F^{\mu\nu}$ are the antisymmetric field tensors for $\omega^{\mu}$, $\rho^{a\mu}$, and $A^{\mu}$, respectively. In the RMF approximation, the meson fields are treated as classical fields and they are replaced by expectation values.

In the STF approximation, the nucleon distribution function can be obtained self-consistently, which is given by
\begin{equation}
\label{eq:nirmf}
 n_{i}(r)=\frac{1}{\pi^2}
       \int_0^{\infty} dk\,k^2\,\left[f_{i}^{k}(r)-f_{\bar{i}}^{k}(r)\right],
\end{equation}
where $f_{i}^{k}$ and $f_{\bar{i}}^{k}$ ($i=p$, $n$) are the occupation probabilities of the particle and antiparticle for momentum $k$. At a finite temperature, the occupation probability is obtained by the Fermi--Dirac distribution. In the PTF approximation, the nucleon distribution function is assumed to have a certain form\rf{4,8}.

At given temperature $T$, proton fraction $Y_p$, and baryon density $\rho_B$, we determine the thermodynamically favored state by minimizing the free energy $F$ with respect to the Wigner--Seitz cell radius. The equilibrium state is the one having the lowest $F$ among droplet, bubble, and homogeneous phases.

\section{Results and discussion}
\label{sec:3}

\begin{figure}[htbp]
\begin{minipage}[t]{0.48\linewidth}
\centering
\includegraphics[bb=0 350 540 775, width=4.5 cm,clip]{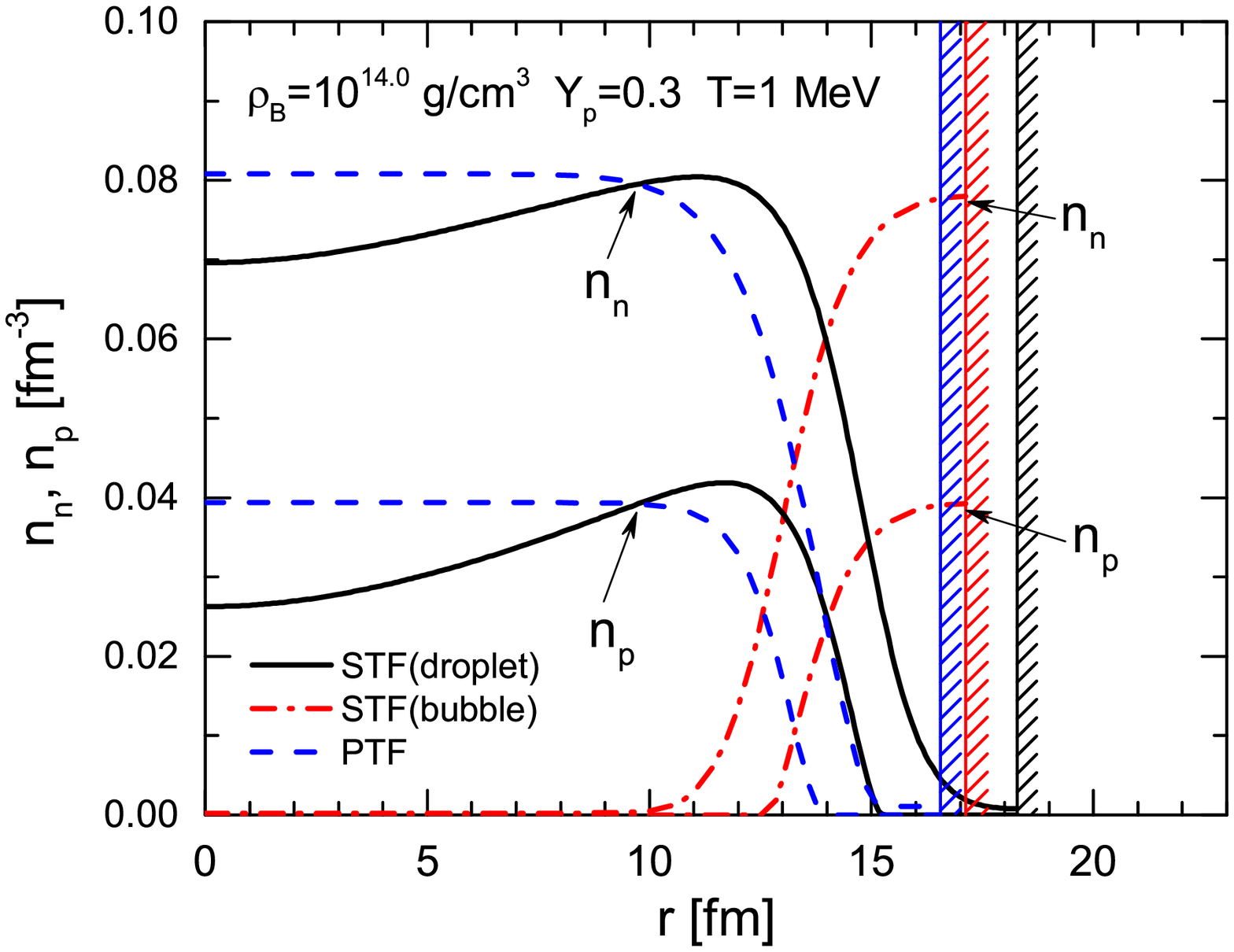}
\caption{\scriptsize{Density distributions of protons and neutrons inside the
Wigner--Seitz cell for the cases of $Y_p=0.3$ and
$\rho_B = 10^{14.0}\, \rm{g\,cm^{-3}}$ at
$T=1$ MeV.
The cell radius is indicated by the hatching.\label{fig:1}}}
\end{minipage}
\hfill
\begin{minipage}[t]{0.48\linewidth}
\centering
\includegraphics[bb=10 330 550 770, width=4.5 cm,clip]{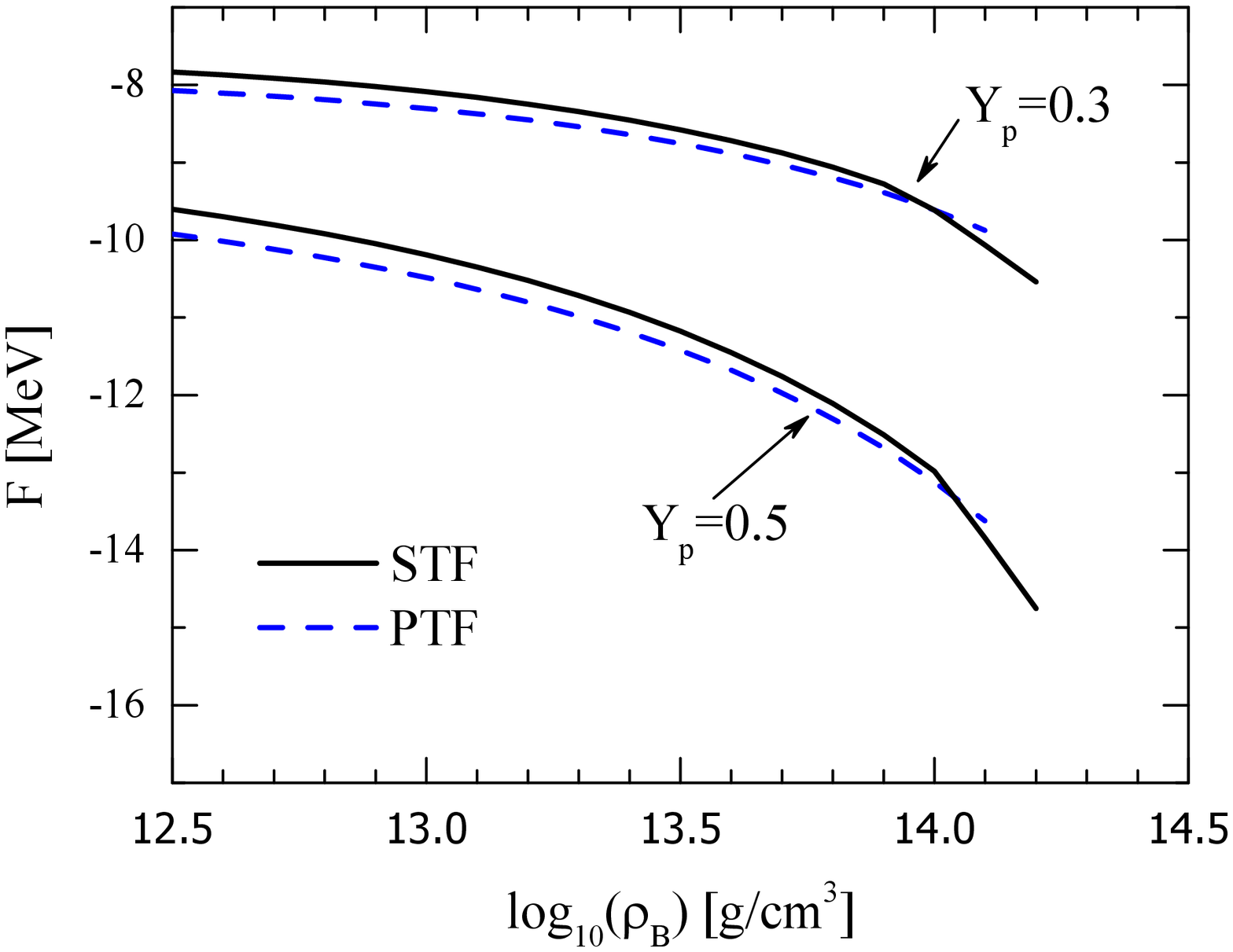}
\caption{\scriptsize{Free energy per baryon $F$ versus
$\rho_B$ for $Y_p=0.3$ and $0.5$ at $T=1$ MeV.
The results of STF (solid lines)
are compared with those of PTF (dashed lines).\label{fig:2}}}
\end{minipage}
\end{figure}

We show in Fig.~\ref{fig:1} the nucleon density distributions in the Wigner--Seitz cell. It is obvious that the densities at the center of the cell are lower than that of the surface region in the STF approximation (solid lines). This is because the protons inside the nucleus are pushed off to the surface by the Coulomb interaction. On the other hand, the nucleon distributions in the PTF approximation (dashed lines) are forced to have a certain form.

We present the Free energy per baryon $F$ versus $\rho_B$ in Fig.~\ref{fig:2}. It is shown that there is a significant decrease for the solid lines at about $\rho_B \sim 10^{13.9}\,\rm{g\,cm^{-3}}$. This is because the bubble phase has a lower free energy than the droplet phase at this density in the STF approximation. As a result, the appearance of the bubble phase can delay the transition to uniform matter. We find that there is a small difference in $F$ between STF and PTF, which may be due to the different treatment of surface effect and nucleon distribution between these two methods.
The value of $F$ in PTF is generally lower than that in STF. This is mainly because the surface energy in PTF is smaller than that in STF.
We note that the surface energy is self-consistently computed in STF, while it is approximately calculated in PTF by using Equation (26) of Ref.[4] with an additional parameter $F_0$. The choice of $F_0=70 \, \rm{MeV\,fm^5}$ in PTF yields smaller surface energies than those of STF, and therefore, smaller free energies are obtained in PTF as shown in Fig.~\ref{fig:2}. In our previous work\cite{9}, the influence of the parameter $F_0$ was examined, and the results of PTF with $F_0=90 \, \rm{MeV\,fm^5}$ could be very close to the values of STF. Considering the wide range of thermodynamic conditions in the whole EOS\cite{4}, the differences between STF and PTF are thought to be negligible and cannot affect the general behavior of the EOS.

\begin{table}[htb]
\scriptsize
\caption{\scriptsize{
Comparison between different methods for the
cases of $Y_p=0.3$ and $T=1$ MeV
at $\rho_B = 10^{13.0}\, \rm{g\,cm^{-3}}$.\label{tab:1}}}
\begin{center}
\begin{tabular}{lccccccc}
\hline\hline
  method & $F$ (MeV)   & $E$ (MeV) & $S$ ($k_B$)   & $E_b$ (MeV) & $E_g$ (MeV) & $E_C$ (MeV) & $R_c$ (fm)\\
 \hline
 \rule{0pt}{12pt}
  STF            & -8.087 & -7.807 & 0.280 & -10.135 & 1.164 & 1.164 & 20.0\\
       PTF ($F_0=70$) & -8.304 & -8.025 & 0.278 & -10.161 & 1.068 & 1.068 & 19.3\\
 \vspace{3pt}
       PTF ($F_0=90$) & -8.023 & -7.748 & 0.275 & -10.080 & 1.166 & 1.166 & 20.3\\
      \hline\hline
\end{tabular}
\end{center}
\end{table}

\end{document}